\documentclass[prb,twocolumn,superscriptaddress,floatfix,shortbibliography]{revtex4-2}
\usepackage{graphicx}
\usepackage{bm}
\usepackage{color,soul}
\usepackage{amsmath,amssymb,amsfonts,float,graphics,epsfig,epstopdf,color,verbatim,tabularx,bm,multirow,appendix,hyperref}

\begin{document}
\title{Low-temperature specific-heat studies on two square-kagome antiferromagnets}
\author{Bo Liu}
\affiliation{Beijing National Laboratory for Condensed Matter Physics, Institute of Physics, Chinese Academy of Sciences, Beijing 100190, China}
\affiliation{School of Physical Sciences, University of Chinese Academy of Sciences, Beijing 100190, China}
\author{Zhenyuan Zeng}
\affiliation{Beijing National Laboratory for Condensed Matter Physics, Institute of Physics, Chinese Academy of Sciences, Beijing 100190, China}
\affiliation{School of Physical Sciences, University of Chinese Academy of Sciences, Beijing 100190, China}
\author{Aini Xu}
\affiliation{Beijing National Laboratory for Condensed Matter Physics, Institute of Physics, Chinese Academy of Sciences, Beijing 100190, China}
\affiliation{School of Physical Sciences, University of Chinese Academy of Sciences, Beijing 100190, China}
\author{Yili Sun}
\affiliation{Beijing National Laboratory for Condensed Matter Physics, Institute of Physics, Chinese Academy of Sciences, Beijing 100190, China}
\affiliation{School of Physical Sciences, University of Chinese Academy of Sciences, Beijing 100190, China}
\author{Olga Yakubovich}
\affiliation{Lomonosov Moscow State University, Moscow 119991, Russia}
\author{Larisa Shvanskaya}
\affiliation{Lomonosov Moscow State University, Moscow 119991, Russia}
\affiliation{National University of Science and Technology “MISiS”, Moscow 119049, Russia}
\author{Shiliang Li}
\email{slli@iphy.ac.cn}
\affiliation{Beijing National Laboratory for Condensed Matter Physics, Institute of Physics, Chinese Academy of Sciences, Beijing 100190, China}
\affiliation{School of Physical Sciences, University of Chinese Academy of Sciences, Beijing 100190, China}
\affiliation{Songshan Lake Materials Laboratory , Dongguan, Guangdong 523808, China}
\author{Alexander Vasiliev}
\email{anvas2000@yahoo.com}
\affiliation{Lomonosov Moscow State University, Moscow 119991, Russia}
\affiliation{National University of Science and Technology “MISiS”, Moscow 119049, Russia}
\begin{abstract}
We studied the low-temperature specific heats of two antiferromagnets with the two-dimensional square-kagome structure, i.e., KCu$_6$AlBiO$_4$(SO$_4$)$_5$Cl (KCu6) and Na$_6$Cu$_7$BiO$_4$(PO$_4$)$_4$[Cl,(OH)]$_3$ (NaCu7) with the structural differences that there are interlayer Cu$^{2+}$ ions in NaCu7. Both materials show no magnetic ordering down to 50 mK. At zero field, the $C/T$ of KCu6 has a finite value when the temperature is close to zero K. Under the magnetic field, a seemingly $T^2$ dependence appears and its coefficient is progressively suppressed by the field. For NaCu7, the specific heat exhibits the $T^2$ dependence at zero field and under fields. The ratio of the quadratic coefficients of KCu6 and NaCu7 at high fields is inversely proportional to ratio of the squared Weiss temperatures, which indicates these two compounds host the same ground state under fields. Our results suggest that the interlayer Cu$^{2+}$ ions in NaCu7 play a negligible role in determination of its ground state. We discuss the possible quantum-spin-liquid states in these compounds and further directions to pursue based our results.
\end{abstract}


\maketitle
\section{introduction}

Geometrical frustrated magnetic materials have long been thought to be the playgrounds for searching quantum spin liquids (QSLs) \cite{RamirezAP94}. The triangle-based lattice structures have drawn particular attentions as a triangle is one of the basic building blocks to produce frustrations.  One of the most popular examples is the two-dimensional (2D) kagome lattice \cite{BalentsL10,NormanMR16,SavaryL17,ZhouY17,BroholmC20}, which consists of corner-sharing triangles with the smallest loop of six sites. Experimentally, herbertsmithite and its related compounds with the kagome structure show strong signatures for the existence of QSLs although there are still debates on the exact nature of their ground states \cite{ShoresMP05,HanTH12,FuM15,HanTH16,KhuntiaP20,FengZL17,WeiY17,FengZL18b,WeiY21,YingFu2021,ZengZ21}. Starting from the kagome lattice, it has been proposed that the  shuriken or square-kagome lattice (SKL) may also host exotic ground states including QSLs \cite{SiddharthanR01,RichterJ09,RousochatzakisI13,NakanoH13,RalkoA15,MoritaK18,LuganT19,AstrakhantsevN21,RichterJ22}. The triangles are also corner-sharing as those in the kagome lattice, but the SKL has smallest loops with four and eight sites. Moreover, the square loops of the SKL endure stronger quantum fluctuations and thus the system exhibits long-range correlations of virtual singlets \cite{RalkoA15}. The ground states of the SKL could be incommensurate orders, pinwheel valence bond crystal (VBC), U(1) and topological QSLs \cite{LuganT19,AstrakhantsevN21}. 

\begin{figure}[tbp]
\includegraphics[width=\columnwidth]{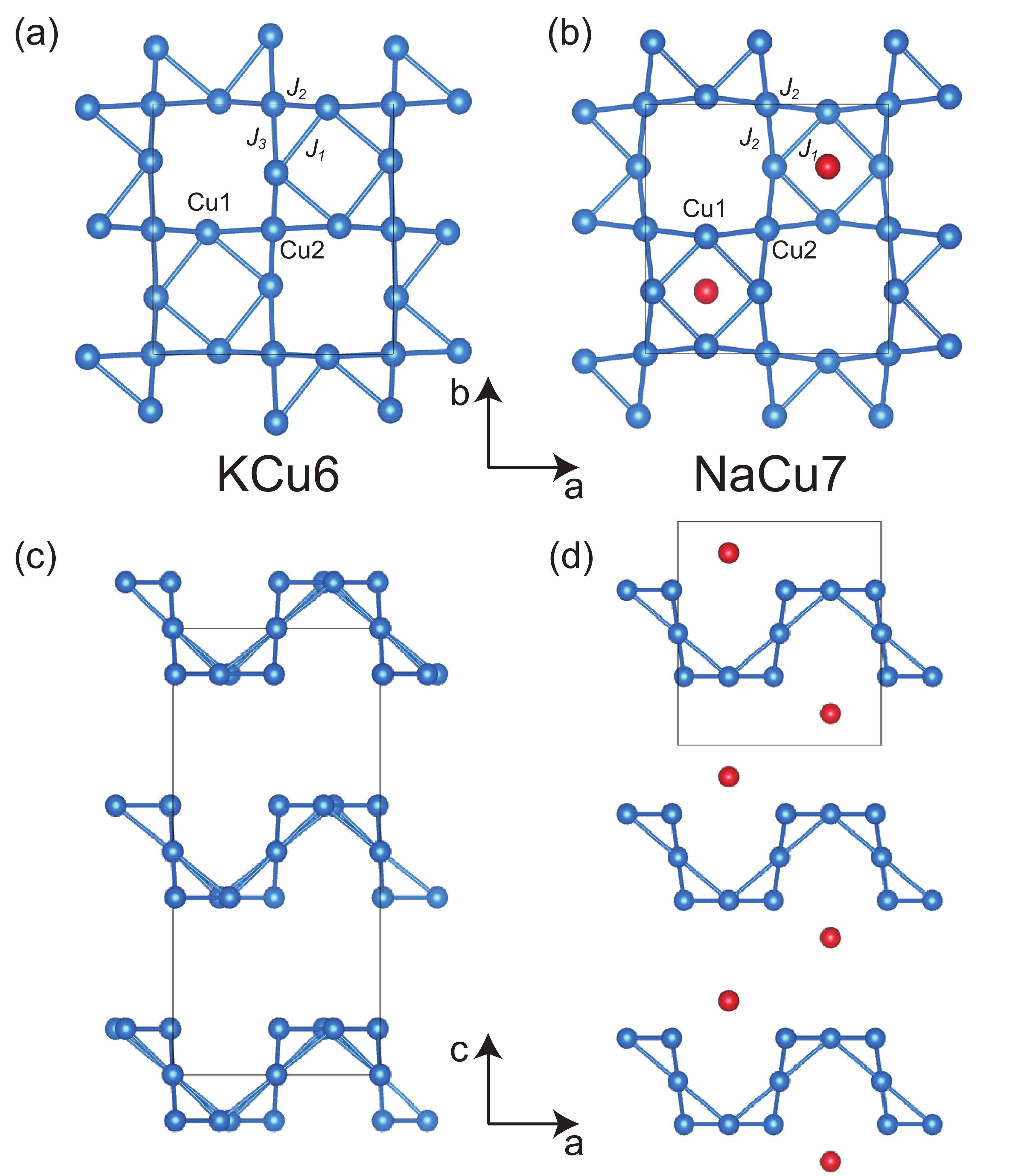}
 \caption{Structures of Cu ions in (a) and (c) KCu6  (Top and side views), and (b) and (d) NaCu7 (Top and side views). The interlayer Cu$^{2+}$ ions in NaCu7 are shown as red balls. The configurations of superexchanges for KCu6 and NCu7 are in (a) and (b), respectively. The black solid lines indicate unit cells. 
}
\label{fig1}
\end{figure}

The material realization of the SKL was only achieved recently in KCu$_6$AlBiO$_4$(SO$_4$)$_5$ (KCu6) \cite{FujihalaM20} and Na$_6$Cu$_7$BiO$_4$(PO$_4$)$_4$[Cl,(OH)]$_3$ (NaCu7) \cite{YakubovichOV21}. Extensive studies have been made on KCu6 down to 50 mK and shown that it may host a gapless QSL ground state \cite{FujihalaM20}. Magnetic-susceptibility and specific-heat measurements on NaCu6 down to 2 K also suggest it may be a promising candidate for the QSLs. We show the structures of Cu$^{2+}$ ions in KCu6 and NaCu7 in Fig. \ref{fig1}, where the SKL structure can be clearly seen from the top view. The most significant difference between the magnetic systems of these two structures is that there are interlayer Cu$^{2+}$ ions for NaCu7. It is known that for kagome magnetic systems, interlayer Cu$^{2+}$ ions act as magnetic impurities and can significantly affect the system properties \cite{HanTH16,VriesMA08,FreedmanDE10,YYHuang2021,KermarrecE11,LIY14,FengZL17,WeiY17,FengZL18b,YingFu2021}. While it seems that the interlayer Cu$^{2+}$ ions in NaCu7 are magnetically isolated from the SKL layers, their effects are only studied down to 2 K \cite{YakubovichOV21}, which makes it unclear whether similar impurity issues also exist in NaCu7. Another important difference between these two compounds is that the Cu1 in KCu6 is not in a symmetrical position so there are two different superexchanges between Cu1 and Cu2, as shown in Fig. \ref{fig1}(a), which means that a minimal $J_1-J_2-J_3$ SKL model is necessary \cite{FujihalaM20}. On the contrary, there is only one $J_2$ between Cu1 and Cu2 in NaCu7 as shown in Fig. \ref{fig1}(b), which probably makes the spin system of NaCu7 more close to the proposed SKL model \cite{SiddharthanR01,RichterJ09,RousochatzakisI13,NakanoH13,RalkoA15,MoritaK18,LuganT19,AstrakhantsevN21}. These differences make it necessary to compare the properties of KCu6 and NaCu7.

In this work, we studied the low-temperature specific heats of KCu6 and NaCu7. We confirm that none of them shows magnetic ordering down to 50 mK. We find that magnetic impurities are few in both compounds. The specific heat of NaCu7 shows a quadratic temperature dependence, whose coefficient linearly decreases with increasing field. For KCu6, the specific heat exhibits almost linear temperature dependence at zero field but quickly changes to the $T^2$ dependence under fields. Interestingly, their specific heats at high fields are very similar and may be directly associated with the values of superexchanges.  Our results suggest that both compounds could host QSL ground states predicted by the SKL model but further studies are needed to solve some issues.

\section{experiments}

Polycrystals of KCu6 and NCu7 were grown by the solid-state reaction and hydrothermal method, respectively, as reported previously \cite{FujihalaM20,YakubovichOV21}. The crystal structures and magnetization were confirmed by the powder x-ray diffraction and a superconducting quantum interference device (SQUID) magnetometer (MPMS 3, Quantum Design), respectively. The specific heats were measured on a Physical Property Measurement System (PPMS, Quantum Design) with the dilution refrigerator option. 

\section{results and discussions}

\begin{figure}[tbp]
\includegraphics[width=\columnwidth]{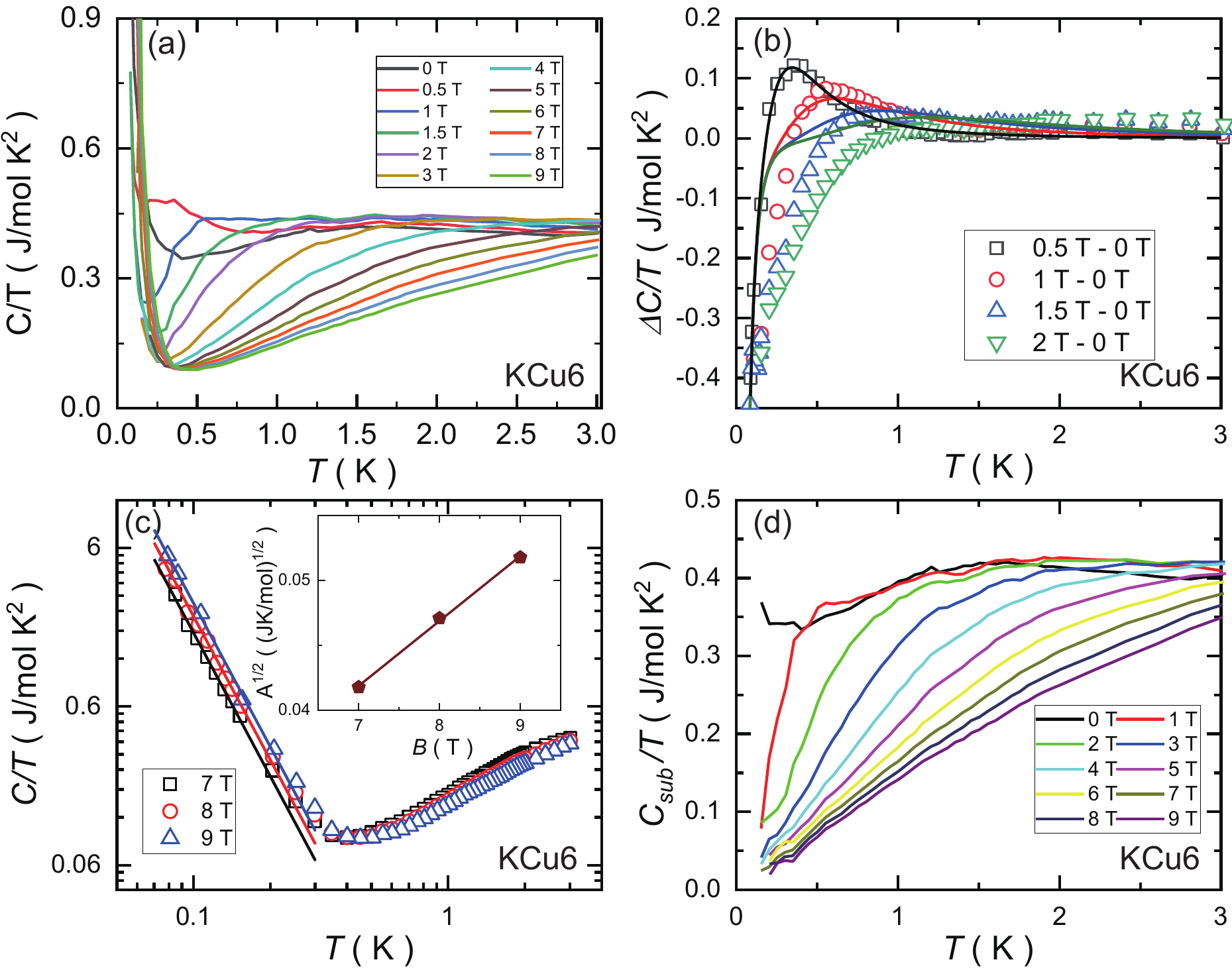}
 \caption{(a) The temperature dependence of $C/T$ of KCu6 under different fields. (b) Field difference of $C/T$ below 2 T as a function of  temperature. The solid line for 0.5T-0T is fitted the result by Eq. (\ref{FieldSchottky}), while other lines are calculated results based on Eq. (\ref{FieldSchottky}). (c) The $C/T$ above 7 T as a function of temperature in the log-log scale. The solid lines are the fitted results by the nuclear Schottky anomaly function as described in the text. The inset shows the field dependence of the fitted value of $A^{1/2}$. The solid line is the linear fitting result. (d) The temperature dependence of $C/T$ of KCu6 under different fields with the impurity and nuclear contributions subtracted. 
}
\label{fig2}
\end{figure}

Figure \ref{fig2}(a) shows the temperature dependence of $C/T$ of  KCu6. Consistent with previous report \cite{FujihalaM20}, there is no magnetic order down to 50 mK.  In analyzing the low-temperature specific heat of a QSL candidate, we typically need to remove two kinds of contributions that are not intrinsic to the magnetic system. The first one is from orphan or weakly correlated magnetic impurities, which are usually weakly correlated so can be described by the Schottky anomaly \cite{VriesMA08}. The specific heat from the Schottky anomaly takes the form $C_{Sch} = NR(\Delta/T)^2e^{\Delta/T}/[1+e^{\Delta/T}]^2$, where $N$ is the impurity concentration, $R$ is the gas constant, and $\Delta = \Delta_0 + g\mu_B H$ is the energy level. Assuming that the change of the whole specific heat mainly comes from these magnetic impurities, we have
\begin{equation}
\Delta C = C(H) - C(0) \approx C_{Sch}(H) - C_{Sch}(0).
\label{FieldSchottky}
\end{equation}
\noindent As shown in Fig. \ref{fig2}(b), this equation can nicely describe the 0.5T-0T data with $g$ = 2.7, $\Delta_0$ = 0.132 K and $N$ = 0.0133, which suggest that there are about 1.33\% orphan magnetic impurities per molecular formula. With these fitted parameters, we can calculate the contribution of magnetic impurities based on Eq. (\ref{FieldSchottky}). With increasing field, the deviation from the data becomes larger, which indicates that there is intrinsic change of the specific heat from the magnetic system. The second non-intrinsic contribution to the specific heat is the nuclear Schottky anomaly, whose energy levels are very small so we only observe its high-temperature tail as $C = A/T^2$. As shown in Fig. \ref{fig2}(c), this function can well describe the data below 2 K for the field larger than 7 T. The inset of Fig. \ref{fig2}(c) shows $A^{1/2}$ as the function of field, which shows a linear field dependence. Accordingly, the contribution from the nuclear Schottky anomaly can be calculated based on the fitted value of $A$.

Figure \ref{fig2}(d) shows $C/T$ of KCu6 with the contributions from both magnetic impurities and nuclear Schottky anomaly subtracted as described above. We note that the phonon part of the specific can be neglected since its $C/T$ at 3 K would be just about 0.03 J/mol K$^2$ with the Debye temperature of 100 K, which should be a very conservative estimate. At 0 T, there seems to be a residual $C/T$ at 0 K, larger than 0.3 J/mol K$^2$. With field applied, the low-temperature $C/T$ is quickly suppressed and shows almost linear temperature dependence at high fields. In other words, the specific heat probably has a quadratic temperature dependence.

\begin{figure}[tbp]
\includegraphics[width=\columnwidth]{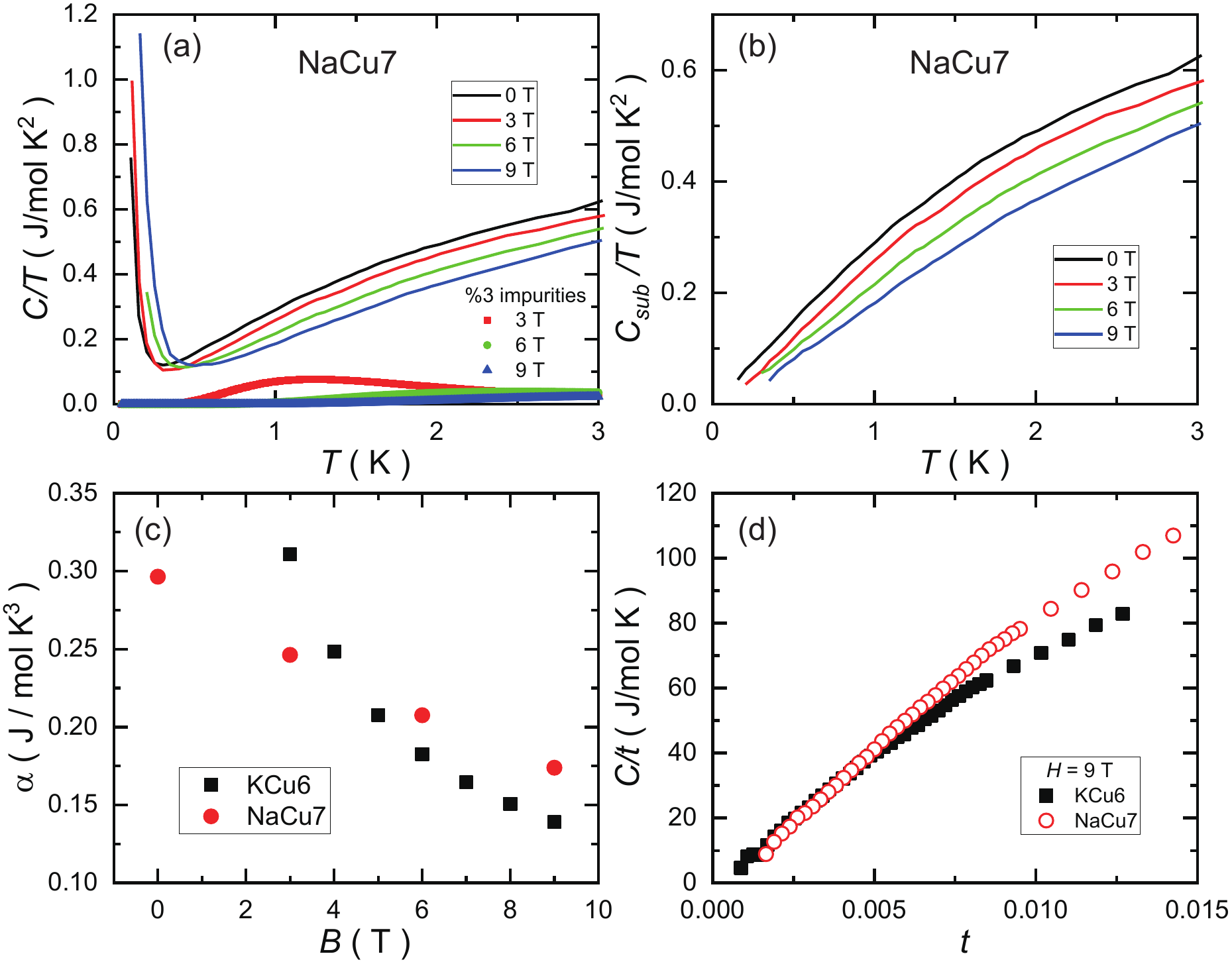}
 \caption{(a) The temperature dependence of $C/T$ of NaCu7 at several fields (lines). The symbols are calculated specific heat for 3\% of magnetic impurities per molecular formula with $g$ = 2 and $\Delta_0$ = 0. (b) The temperature dependence of $C/T$ with the nuclear contributions subtracted. (c) The field dependence of $\alpha$ for KCu6 and NaCu7, which is defined in the main text. (d) The $t$ dependence of $C/t$ for KCu6 and NaCu7 at 9 T. Here $t$ is the temperature normalized by the Weiss temperature as defined in the main text.
}
\label{fig3}
\end{figure}

The low-temperature $C/T$ of NaCu7 are shown in Fig. \ref{fig3}(a). The nuclear Schottky anomaly still clearly presents at very low temperature. After removing the nuclear Schottky anomaly, all $C/T$ tend to linearly decrease to zero, as shown in Fig. \ref{fig3}(b). The specific heat from the magnetic system is also progressively suppressed by the magnetic field, which cannot be explained by the existence of orphan magnetic impurities in this system. To see this, we can calculate the specific heat of 3\% magnetic impurities per molecular formula by using the Schottky anomaly function, as shown in Fig. \ref{fig3}(a). The field dependence of the calculated values clearly suggests that if there are any weakly correlated magnetic impurities, their content should be negligible and thus the field dependence of the measured specific heat is from the SKL magnetic system. It is particularly interesting that the $C/T$ at 9 T in NaCu7 looks to be very similar to that in KCu6. We can fit the specific heat of both KCu6 and NaCu7 below certain temperatures by the function $C/T = A/T^3 + \alpha T$.  The field dependence of $\alpha$ is shown in Fig. \ref{fig3}(c). For NaCu7, $\alpha$ linearly depends on the magnetic field, whereas it increases quickly with decreasing field below about 6 T for KCu6. 

Our low-temperature specific-heat results provide some key information on the  square-kagome magnets. The first important result is that the existence of interlayer Cu$^{2+}$ ions in NaCu7 does not result in any magnetic ordering. This result suggests that unlike the well-known herbertsmithite and its siblings \cite{HanTH16,VriesMA08,FreedmanDE10,YYHuang2021,KermarrecE11,LIY14,FengZL17,WeiY17,FengZL18b,YingFu2021}, the interlayer Cu$^{2+}$ ions in the SKL materials may not significantly affect the ground states. This is a good news for further studies on these materials. While this may be understood that the interlayer Cu$^{2+}$ ions are probably magnetically isolated from the SKL layers, our specific-heat measurements suggest that they do not act as orphan or weakly correlated spins. Further studies are needed to see whether the spins of the interlayer Cu$^{2+}$ magnetic system themselves are strongly correlated or they are not independent of the SKL layers.

The second important result is that the specific heats of KCu6 are NaCu7 are very similar at high magnetic fields. As shown above, the specific heat at high fields of both materials roughly shows $T^2$ dependence at low temperatures. The ratio of the coefficient $\alpha$ for the quadratic term between NaCu7 and KCu6 is about 1.27 $-$ 1.33 at 9 T depending on the fitting ranges. The Weiss temperatures $\Theta_{CW}$ for NaCu7 and KCu6 are -212 and -237 K, respectively \cite{FujihalaM20,YakubovichOV21}. Assuming that the superexchange $J$ is proportional to the Weiss temperature, this suggests that the ratio of $\alpha$ may be related to the ratio of $1/J^2$, which is about 1.25. We thus define the normalized temperature $t = T/|\Theta_{CW}|$ and plot $C/t$ vs $t$ in Fig. \ref{fig3}(d). It is clear that the data of KCu6 and NaCu7 are almost overlapped with each other below $t \sim$ 0.005, which is about 1 K. 

Our results suggest that both KCu6 and NaCu7 are good candidates for QSLs. None of them shows magnetic ordering down to 50 mK despite of their large Weiss temperatures. In the optimistic view, both of them can host QSL ground states. At zero field, the $C/T$ of KCu6 seems to have a finite value at zero K, which indicates a linear temperature dependence of $C$ that could come from spinon Fermi surfaces. At high fields, the specific heat shows a $T^2$ temperature dependence. For NaCu7, the specific heat always exhibits the $T^2$ temperature dependence. This quadratic temperature dependence of the specific heat for QSLs has been shown to exist in the U(1) Dirac QSL, whose coefficient is proportional to $1/J^2$ \cite{RanY07}. Interestingly, the SKL model could indeed host a U(1) Dirac spin liquid \cite{AstrakhantsevN21,XuXY19}. It should be pointed it out that although our data indicate gapless states for both compounds, we cannot rule out that very small gaps may exist, which are covered by the large nuclear Schottky anomaly at low temperatures.

Of course, there exist few potiential problems to be clarified before one confirm the above note of optimism. First, the low-temperature specific heat of the U(1) Dirac QSLs should have linear-T component under magnetic field and increase with the increasing field due to the formation of spinon Fermi pockets \cite{RanY07,ZengZ21}. However, for both KCu6 and NaCu7, the low-temperature specific heats are suppressed by the field. We also note that for a trivial 2D magnetism, the low-temperature specific heat could also exhibit a $T^2$ temperature dependence with the coefficient proportional to $1/J^2$ \cite{TakahashiM89,LiuBG90}. Second, it is rather hard to understand a crossover or transition from a spinon Fermi-surface state to a U(1) Dirac state with such small change of the magnetic field in KCu6. One may also expect some signatures for this kind of crossover or transition between two distinct ground states, which cannot be identified in our measurements. Third, the different configuration of superexchanges in these two compounds, as shown in Fig. 1, seems to have no effects on the nature of the ground states under fields. If assuming that $J_1$ is proportional to the Cu1-O-Cu1 bond angle, we can estimate that the ratio of $J_1^2$ between KCu6 and NaCu7 is about 1.3 according to previous results on the 2D copper oxides compounds \cite{ShimizuT03,RocquefelteX12}. This value is consistent with the normalized factor in Fig. \ref{fig3}(d) and seems to suggest that the low-energy excitations in the SKL model are probably mainly associated with $J_1$ since we cannot find such coincidence if $J_2$ and $J_3$ are also considered. If this is true, the finite value of $C/T$ at 0 K in KCu6 may be related to the different values of $J_2$ and $J_3$, especially the large value of $J_2$.  Of course, this is rather a naive hypothesis. All these issues need to be investigated further both experimentally and theoretically.

\section{conclusions}
Our studies on the low-temperature specific heats of KCu6 and NaCu7 provide some basic key information that are important to further study their magnetic ground states. We show that magnetic impurities are few in both compounds and the existence of interlayer Cu$^{2+}$ ions probably has negligible effects on the magnetic system of the SKL layers. The zero-field specific heat of KCu6 may consist of a large linear temperature dependence, which is easily suppressed by the magnetic field.  The low-temperature specific heat of KCu6 under magnetic fields and that of NaCu7 are very similar, which suggests that both of them may be good platforms to study the SKL model. 

\begin{acknowledgments}

This work is supported by the National Key Research and Development Program of China (Grants No. 2021YFA1400401, No. 2017YFA0302903), the National Natural Science Foundation of China (Grants No. 11961160699, No. 11874401), the Strategic Priority Research Program(B) of the Chinese Academy of Sciences (Grants No. XDB33000000, No. GJTD-2020-01), and the Megagrant program of the Government of Russian Federation (Grant No. 075-15-2021-604).

\end{acknowledgments}


\begin{thebibliography}{39}%
\makeatletter
\providecommand \@ifxundefined [1]{%
 \@ifx{#1\undefined}
}%
\providecommand \@ifnum [1]{%
 \ifnum #1\expandafter \@firstoftwo
 \else \expandafter \@secondoftwo
 \fi
}%
\providecommand \@ifx [1]{%
 \ifx #1\expandafter \@firstoftwo
 \else \expandafter \@secondoftwo
 \fi
}%
\providecommand \natexlab [1]{#1}%
\providecommand \enquote  [1]{``#1''}%
\providecommand \bibnamefont  [1]{#1}%
\providecommand \bibfnamefont [1]{#1}%
\providecommand \citenamefont [1]{#1}%
\providecommand \href@noop [0]{\@secondoftwo}%
\providecommand \href [0]{\begingroup \@sanitize@url \@href}%
\providecommand \@href[1]{\@@startlink{#1}\@@href}%
\providecommand \@@href[1]{\endgroup#1\@@endlink}%
\providecommand \@sanitize@url [0]{\catcode `\\12\catcode `\$12\catcode
  `\&12\catcode `\#12\catcode `\^12\catcode `\_12\catcode `\%12\relax}%
\providecommand \@@startlink[1]{}%
\providecommand \@@endlink[0]{}%
\providecommand \url  [0]{\begingroup\@sanitize@url \@url }%
\providecommand \@url [1]{\endgroup\@href {#1}{\urlprefix }}%
\providecommand \urlprefix  [0]{URL }%
\providecommand \Eprint [0]{\href }%
\providecommand \doibase [0]{https://doi.org/}%
\providecommand \selectlanguage [0]{\@gobble}%
\providecommand \bibinfo  [0]{\@secondoftwo}%
\providecommand \bibfield  [0]{\@secondoftwo}%
\providecommand \translation [1]{[#1]}%
\providecommand \BibitemOpen [0]{}%
\providecommand \bibitemStop [0]{}%
\providecommand \bibitemNoStop [0]{.\EOS\space}%
\providecommand \EOS [0]{\spacefactor3000\relax}%
\providecommand \BibitemShut  [1]{\csname bibitem#1\endcsname}%
\let\auto@bib@innerbib\@empty
\bibitem [{\citenamefont {Ramirez}(1994)}]{RamirezAP94}%
  \BibitemOpen
  \bibfield  {author} {\bibinfo {author} {\bibfnamefont {A.~P.}\ \bibnamefont
  {Ramirez}},\ }\bibfield  {title} {\bibinfo {title} {Strongly geometrically
  frustrated magnets},\ }\href
  {https://doi.org/10.1146/annurev.ms.24.080194.002321} {\bibfield  {journal}
  {\bibinfo  {journal} {Annu. Rev. Mater. Res.}\ }\textbf {\bibinfo {volume}
  {24}},\ \bibinfo {pages} {453} (\bibinfo {year} {1994})}\BibitemShut
  {NoStop}%
\bibitem [{\citenamefont {Balents}(2010)}]{BalentsL10}%
  \BibitemOpen
  \bibfield  {author} {\bibinfo {author} {\bibfnamefont {L.}~\bibnamefont
  {Balents}},\ }\bibfield  {title} {\bibinfo {title} {Spin liquids in
  frustrated magnets},\ }\href {https://doi.org/10.1038/nature08917} {\bibfield
   {journal} {\bibinfo  {journal} {Nature}\ }\textbf {\bibinfo {volume}
  {464}},\ \bibinfo {pages} {199} (\bibinfo {year} {2010})}\BibitemShut
  {NoStop}%
\bibitem [{\citenamefont {Norman}(2016)}]{NormanMR16}%
  \BibitemOpen
  \bibfield  {author} {\bibinfo {author} {\bibfnamefont {M.~R.}\ \bibnamefont
  {Norman}},\ }\bibfield  {title} {\bibinfo {title} {Colloquium:
  Herbertsmithite and the search for the quantum spin liquid},\ }\href
  {https://doi.org/10.1103/RevModPhys.88.041002} {\bibfield  {journal}
  {\bibinfo  {journal} {Rev. Mod. Phys.}\ }\textbf {\bibinfo {volume} {88}},\
  \bibinfo {pages} {041002} (\bibinfo {year} {2016})}\BibitemShut {NoStop}%
\bibitem [{\citenamefont {Savary}\ and\ \citenamefont
  {Balents}(2017)}]{SavaryL17}%
  \BibitemOpen
  \bibfield  {author} {\bibinfo {author} {\bibfnamefont {L.}~\bibnamefont
  {Savary}}\ and\ \bibinfo {author} {\bibfnamefont {L.}~\bibnamefont
  {Balents}},\ }\bibfield  {title} {\bibinfo {title} {Quantum spin liquids: {A}
  review},\ }\href {http://dx.doi.org/10.1088/0034-4885/80/1/016502} {\bibfield
   {journal} {\bibinfo  {journal} {Rep. Prog. Phys.}\ }\textbf {\bibinfo
  {volume} {80}},\ \bibinfo {pages} {016502} (\bibinfo {year}
  {2017})}\BibitemShut {NoStop}%
\bibitem [{\citenamefont {Zhou}\ \emph {et~al.}(2017)\citenamefont {Zhou},
  \citenamefont {Kanoda},\ and\ \citenamefont {Ng}}]{ZhouY17}%
  \BibitemOpen
  \bibfield  {author} {\bibinfo {author} {\bibfnamefont {Y.}~\bibnamefont
  {Zhou}}, \bibinfo {author} {\bibfnamefont {K.}~\bibnamefont {Kanoda}},\ and\
  \bibinfo {author} {\bibfnamefont {T.-K.}\ \bibnamefont {Ng}},\ }\bibfield
  {title} {\bibinfo {title} {Quantum spin liquid states},\ }\href
  {https://doi.org/10.1103/RevModPhys.89.025003} {\bibfield  {journal}
  {\bibinfo  {journal} {Rev. Mod. Phys.}\ }\textbf {\bibinfo {volume} {89}},\
  \bibinfo {pages} {025003} (\bibinfo {year} {2017})}\BibitemShut {NoStop}%
\bibitem [{\citenamefont {BROHOLM}\ \emph {et~al.}(2020)\citenamefont
  {BROHOLM}, \citenamefont {CAVA}, \citenamefont {KIVELSON}, \citenamefont
  {NOCERA}, \citenamefont {NORMAN},\ and\ \citenamefont
  {SENTHIL}}]{BroholmC20}%
  \BibitemOpen
  \bibfield  {author} {\bibinfo {author} {\bibfnamefont {C.}~\bibnamefont
  {BROHOLM}}, \bibinfo {author} {\bibfnamefont {R.~J.}\ \bibnamefont {CAVA}},
  \bibinfo {author} {\bibfnamefont {S.~A.}\ \bibnamefont {KIVELSON}}, \bibinfo
  {author} {\bibfnamefont {D.~G.}\ \bibnamefont {NOCERA}}, \bibinfo {author}
  {\bibfnamefont {M.~R.}\ \bibnamefont {NORMAN}},\ and\ \bibinfo {author}
  {\bibfnamefont {T.}~\bibnamefont {SENTHIL}},\ }\bibfield  {title} {\bibinfo
  {title} {Quantum spin liquids},\ }\href
  {https://doi.org/10.1126/science.aay0668} {\bibfield  {journal} {\bibinfo
  {journal} {Science}\ }\textbf {\bibinfo {volume} {367}},\ \bibinfo {pages}
  {eaay0668} (\bibinfo {year} {2020})}\BibitemShut {NoStop}%
\bibitem [{\citenamefont {Shores}\ \emph {et~al.}(2005)\citenamefont {Shores},
  \citenamefont {Nytko}, \citenamefont {Bartlett},\ and\ \citenamefont
  {Nocera}}]{ShoresMP05}%
  \BibitemOpen
  \bibfield  {author} {\bibinfo {author} {\bibfnamefont {M.~P.}\ \bibnamefont
  {Shores}}, \bibinfo {author} {\bibfnamefont {E.~A.}\ \bibnamefont {Nytko}},
  \bibinfo {author} {\bibfnamefont {B.~M.}\ \bibnamefont {Bartlett}},\ and\
  \bibinfo {author} {\bibfnamefont {D.~G.}\ \bibnamefont {Nocera}},\ }\bibfield
   {title} {\bibinfo {title} {A structurally perfect s = 1/2 kagom$\acute{3}$
  antiferromagnet},\ }\href {https://doi.org/10.1021/ja053891p} {\bibfield
  {journal} {\bibinfo  {journal} {J. Am. Chem. Soc.}\ }\textbf {\bibinfo
  {volume} {127}},\ \bibinfo {pages} {13462} (\bibinfo {year}
  {2005})}\BibitemShut {NoStop}%
\bibitem [{\citenamefont {Han}\ \emph {et~al.}(2012)\citenamefont {Han},
  \citenamefont {Helton}, \citenamefont {Chu}, \citenamefont {Nocera},
  \citenamefont {Rodriguez-Rivera}, \citenamefont {Broholm},\ and\
  \citenamefont {Lee}}]{HanTH12}%
  \BibitemOpen
  \bibfield  {author} {\bibinfo {author} {\bibfnamefont {T.~H.}\ \bibnamefont
  {Han}}, \bibinfo {author} {\bibfnamefont {J.~S.}\ \bibnamefont {Helton}},
  \bibinfo {author} {\bibfnamefont {S.}~\bibnamefont {Chu}}, \bibinfo {author}
  {\bibfnamefont {D.~G.}\ \bibnamefont {Nocera}}, \bibinfo {author}
  {\bibfnamefont {J.~A.}\ \bibnamefont {Rodriguez-Rivera}}, \bibinfo {author}
  {\bibfnamefont {C.}~\bibnamefont {Broholm}},\ and\ \bibinfo {author}
  {\bibfnamefont {Y.~S.}\ \bibnamefont {Lee}},\ }\bibfield  {title} {\bibinfo
  {title} {Fractionalized excitations in the spin-liquid state of a
  kagome-lattice antiferromagnet},\ }\href
  {https://doi.org/10.1038/nature11659} {\bibfield  {journal} {\bibinfo
  {journal} {Nature}\ }\textbf {\bibinfo {volume} {492}},\ \bibinfo {pages}
  {406} (\bibinfo {year} {2012})}\BibitemShut {NoStop}%
\bibitem [{\citenamefont {Fu}\ \emph {et~al.}(2015)\citenamefont {Fu},
  \citenamefont {Imai}, \citenamefont {Han},\ and\ \citenamefont
  {Lee}}]{FuM15}%
  \BibitemOpen
  \bibfield  {author} {\bibinfo {author} {\bibfnamefont {M.}~\bibnamefont
  {Fu}}, \bibinfo {author} {\bibfnamefont {T.}~\bibnamefont {Imai}}, \bibinfo
  {author} {\bibfnamefont {T.-H.}\ \bibnamefont {Han}},\ and\ \bibinfo {author}
  {\bibfnamefont {Y.~S.}\ \bibnamefont {Lee}},\ }\bibfield  {title} {\bibinfo
  {title} {Evidence for a gapped spin-liquid ground state in a kagome
  heisenberg antiferromagnet},\ }\href
  {https://doi.org/10.1126/science.aab2120} {\bibfield  {journal} {\bibinfo
  {journal} {Science}\ }\textbf {\bibinfo {volume} {350}},\ \bibinfo {pages}
  {655} (\bibinfo {year} {2015})}\BibitemShut {NoStop}%
\bibitem [{\citenamefont {Han}\ \emph {et~al.}(2016)\citenamefont {Han},
  \citenamefont {Norman}, \citenamefont {Wen}, \citenamefont
  {Rodriguez-Rivera}, \citenamefont {Helton}, \citenamefont {Broholm},\ and\
  \citenamefont {Lee}}]{HanTH16}%
  \BibitemOpen
  \bibfield  {author} {\bibinfo {author} {\bibfnamefont {T.-H.}\ \bibnamefont
  {Han}}, \bibinfo {author} {\bibfnamefont {M.~R.}\ \bibnamefont {Norman}},
  \bibinfo {author} {\bibfnamefont {J.-J.}\ \bibnamefont {Wen}}, \bibinfo
  {author} {\bibfnamefont {J.~A.}\ \bibnamefont {Rodriguez-Rivera}}, \bibinfo
  {author} {\bibfnamefont {J.~S.}\ \bibnamefont {Helton}}, \bibinfo {author}
  {\bibfnamefont {C.}~\bibnamefont {Broholm}},\ and\ \bibinfo {author}
  {\bibfnamefont {Y.~S.}\ \bibnamefont {Lee}},\ }\bibfield  {title} {\bibinfo
  {title} {Correlated impurities and intrinsic spin-liquid physics in the
  kagome material herbertsmithite},\ }\href
  {https://doi.org/10.1103/PhysRevB.94.060409} {\bibfield  {journal} {\bibinfo
  {journal} {Phys. Rev. B}\ }\textbf {\bibinfo {volume} {94}},\ \bibinfo
  {pages} {060409} (\bibinfo {year} {2016})}\BibitemShut {NoStop}%
\bibitem [{\citenamefont {Khuntia}\ \emph {et~al.}(2020)\citenamefont
  {Khuntia}, \citenamefont {Velazquez}, \citenamefont {Barth\'{e}lemy},
  \citenamefont {Bert}, \citenamefont {Kermarrec}, \citenamefont {Legros},
  \citenamefont {Bernu}, \citenamefont {L.~Messio},\ and\ \citenamefont
  {Mendels}}]{KhuntiaP20}%
  \BibitemOpen
  \bibfield  {author} {\bibinfo {author} {\bibfnamefont {P.}~\bibnamefont
  {Khuntia}}, \bibinfo {author} {\bibfnamefont {M.}~\bibnamefont {Velazquez}},
  \bibinfo {author} {\bibfnamefont {Q.}~\bibnamefont {Barth\'{e}lemy}},
  \bibinfo {author} {\bibfnamefont {F.}~\bibnamefont {Bert}}, \bibinfo {author}
  {\bibfnamefont {E.}~\bibnamefont {Kermarrec}}, \bibinfo {author}
  {\bibfnamefont {A.}~\bibnamefont {Legros}}, \bibinfo {author} {\bibfnamefont
  {B.}~\bibnamefont {Bernu}}, \bibinfo {author} {\bibfnamefont {A.~Z.}\
  \bibnamefont {L.~Messio}},\ and\ \bibinfo {author} {\bibfnamefont
  {P.}~\bibnamefont {Mendels}},\ }\bibfield  {title} {\bibinfo {title} {Gapless
  ground state in the archetypal quantum kagome antiferromagnet
  {ZnCu$_3$(OH)$_6$Cl$_2$}},\ }\href
  {https://doi.org/10.1038/s41567-020-0792-1} {\bibfield  {journal} {\bibinfo
  {journal} {Nat. Phys.}\ }\textbf {\bibinfo {volume} {16}},\ \bibinfo {pages}
  {469} (\bibinfo {year} {2020})}\BibitemShut {NoStop}%
\bibitem [{\citenamefont {Feng}\ \emph {et~al.}(2017)\citenamefont {Feng},
  \citenamefont {Li}, \citenamefont {Meng}, \citenamefont {Yi}, \citenamefont
  {Wei}, \citenamefont {Zhang}, \citenamefont {Wang}, \citenamefont {Jiang},
  \citenamefont {Liu}, \citenamefont {Li}, \citenamefont {Liu}, \citenamefont
  {Luo}, \citenamefont {Li}, \citenamefont {qing Zheng}, \citenamefont {Meng},
  \citenamefont {Mei},\ and\ \citenamefont {Shi}}]{FengZL17}%
  \BibitemOpen
  \bibfield  {author} {\bibinfo {author} {\bibfnamefont {Z.}~\bibnamefont
  {Feng}}, \bibinfo {author} {\bibfnamefont {Z.}~\bibnamefont {Li}}, \bibinfo
  {author} {\bibfnamefont {X.}~\bibnamefont {Meng}}, \bibinfo {author}
  {\bibfnamefont {W.}~\bibnamefont {Yi}}, \bibinfo {author} {\bibfnamefont
  {Y.}~\bibnamefont {Wei}}, \bibinfo {author} {\bibfnamefont {J.}~\bibnamefont
  {Zhang}}, \bibinfo {author} {\bibfnamefont {Y.-C.}\ \bibnamefont {Wang}},
  \bibinfo {author} {\bibfnamefont {W.}~\bibnamefont {Jiang}}, \bibinfo
  {author} {\bibfnamefont {Z.}~\bibnamefont {Liu}}, \bibinfo {author}
  {\bibfnamefont {S.}~\bibnamefont {Li}}, \bibinfo {author} {\bibfnamefont
  {F.}~\bibnamefont {Liu}}, \bibinfo {author} {\bibfnamefont {J.}~\bibnamefont
  {Luo}}, \bibinfo {author} {\bibfnamefont {S.}~\bibnamefont {Li}}, \bibinfo
  {author} {\bibfnamefont {G.}~\bibnamefont {qing Zheng}}, \bibinfo {author}
  {\bibfnamefont {Z.~Y.}\ \bibnamefont {Meng}}, \bibinfo {author}
  {\bibfnamefont {J.-W.}\ \bibnamefont {Mei}},\ and\ \bibinfo {author}
  {\bibfnamefont {Y.}~\bibnamefont {Shi}},\ }\bibfield  {title} {\bibinfo
  {title} {Gapped spin-1/2 spinon excitations in a new kagome quantum spin
  liquid compound {Cu$_3$Zn(OH)$_6$FBr}},\ }\href
  {https://doi.org/10.1088/0256-307X/34/7/077502} {\bibfield  {journal}
  {\bibinfo  {journal} {Chin. Phys. Lett.}\ }\textbf {\bibinfo {volume} {34}},\
  \bibinfo {eid} {077502} (\bibinfo {year} {2017})}\BibitemShut {NoStop}%
\bibitem [{\citenamefont {{Wei}}\ \emph {et~al.}(2017)\citenamefont {{Wei}},
  \citenamefont {{Feng}}, \citenamefont {{Lohstroh}}, \citenamefont {{Yu}},
  \citenamefont {{Le}}, \citenamefont {{dela Cruz}}, \citenamefont {{Yi}},
  \citenamefont {{Ding}}, \citenamefont {{Zhang}}, \citenamefont {{Tan}},
  \citenamefont {{Shu}}, \citenamefont {{Wang}}, \citenamefont {{Wu}},
  \citenamefont {{Luo}}, \citenamefont {{Mei}}, \citenamefont {{Yang}},
  \citenamefont {{Sheng}}, \citenamefont {{Li}}, \citenamefont {{Qi}},
  \citenamefont {{Meng}}, \citenamefont {{Shi}},\ and\ \citenamefont
  {{Li}}}]{WeiY17}%
  \BibitemOpen
  \bibfield  {author} {\bibinfo {author} {\bibfnamefont {Y.}~\bibnamefont
  {{Wei}}}, \bibinfo {author} {\bibfnamefont {Z.}~\bibnamefont {{Feng}}},
  \bibinfo {author} {\bibfnamefont {W.}~\bibnamefont {{Lohstroh}}}, \bibinfo
  {author} {\bibfnamefont {D.~H.}\ \bibnamefont {{Yu}}}, \bibinfo {author}
  {\bibfnamefont {D.}~\bibnamefont {{Le}}}, \bibinfo {author} {\bibfnamefont
  {C.}~\bibnamefont {{dela Cruz}}}, \bibinfo {author} {\bibfnamefont
  {W.}~\bibnamefont {{Yi}}}, \bibinfo {author} {\bibfnamefont {Z.~F.}\
  \bibnamefont {{Ding}}}, \bibinfo {author} {\bibfnamefont {J.}~\bibnamefont
  {{Zhang}}}, \bibinfo {author} {\bibfnamefont {C.}~\bibnamefont {{Tan}}},
  \bibinfo {author} {\bibfnamefont {L.}~\bibnamefont {{Shu}}}, \bibinfo
  {author} {\bibfnamefont {Y.-C.}\ \bibnamefont {{Wang}}}, \bibinfo {author}
  {\bibfnamefont {H.-Q.}\ \bibnamefont {{Wu}}}, \bibinfo {author}
  {\bibfnamefont {J.}~\bibnamefont {{Luo}}}, \bibinfo {author} {\bibfnamefont
  {J.-W.}\ \bibnamefont {{Mei}}}, \bibinfo {author} {\bibfnamefont
  {F.}~\bibnamefont {{Yang}}}, \bibinfo {author} {\bibfnamefont {X.-L.}\
  \bibnamefont {{Sheng}}}, \bibinfo {author} {\bibfnamefont {W.}~\bibnamefont
  {{Li}}}, \bibinfo {author} {\bibfnamefont {Y.}~\bibnamefont {{Qi}}}, \bibinfo
  {author} {\bibfnamefont {Z.~Y.}\ \bibnamefont {{Meng}}}, \bibinfo {author}
  {\bibfnamefont {Y.}~\bibnamefont {{Shi}}},\ and\ \bibinfo {author}
  {\bibfnamefont {S.}~\bibnamefont {{Li}}},\ }\bibfield  {title} {\bibinfo
  {title} {{Evidence for the topological order in a kagome antiferromagnet}},\
  }\href@noop {} {\bibfield  {journal} {\bibinfo  {journal} {arXiv e-prints}\
  ,\ \bibinfo {eid} {arXiv:1710.02991}} (\bibinfo {year} {2017})},\ \Eprint
  {https://arxiv.org/abs/1710.02991} {arXiv:1710.02991 [cond-mat.str-el]}
  \BibitemShut {NoStop}%
\bibitem [{\citenamefont {Feng}\ \emph {et~al.}(2018)\citenamefont {Feng},
  \citenamefont {Yi}, \citenamefont {Zhu}, \citenamefont {Wei}, \citenamefont
  {Miao}, \citenamefont {Ma}, \citenamefont {Luo}, \citenamefont {Li},
  \citenamefont {Meng},\ and\ \citenamefont {Shi}}]{FengZL18b}%
  \BibitemOpen
  \bibfield  {author} {\bibinfo {author} {\bibfnamefont {Z.}~\bibnamefont
  {Feng}}, \bibinfo {author} {\bibfnamefont {W.}~\bibnamefont {Yi}}, \bibinfo
  {author} {\bibfnamefont {K.}~\bibnamefont {Zhu}}, \bibinfo {author}
  {\bibfnamefont {Y.}~\bibnamefont {Wei}}, \bibinfo {author} {\bibfnamefont
  {S.}~\bibnamefont {Miao}}, \bibinfo {author} {\bibfnamefont {J.}~\bibnamefont
  {Ma}}, \bibinfo {author} {\bibfnamefont {J.}~\bibnamefont {Luo}}, \bibinfo
  {author} {\bibfnamefont {S.}~\bibnamefont {Li}}, \bibinfo {author}
  {\bibfnamefont {Z.~Y.}\ \bibnamefont {Meng}},\ and\ \bibinfo {author}
  {\bibfnamefont {Y.}~\bibnamefont {Shi}},\ }\bibfield  {title} {\bibinfo
  {title} {From claringbullite to a new spin liquid candidate
  {Cu$_3$Zn(OH)$_6$FCl}},\ }\href
  {https://doi.org/10.1088/0256-307X/36/1/017502} {\bibfield  {journal}
  {\bibinfo  {journal} {Chin. Phys. Lett.}\ }\textbf {\bibinfo {volume} {36}},\
  \bibinfo {pages} {017502} (\bibinfo {year} {2018})}\BibitemShut {NoStop}%
\bibitem [{\citenamefont {Wei}\ \emph {et~al.}(2021)\citenamefont {Wei},
  \citenamefont {Ma}, \citenamefont {Feng}, \citenamefont {Zhang},
  \citenamefont {Zhang}, \citenamefont {Yang}, \citenamefont {Qi},
  \citenamefont {Meng}, \citenamefont {Wang}, \citenamefont {Shi},\ and\
  \citenamefont {Li}}]{WeiY21}%
  \BibitemOpen
  \bibfield  {author} {\bibinfo {author} {\bibfnamefont {Y.}~\bibnamefont
  {Wei}}, \bibinfo {author} {\bibfnamefont {X.}~\bibnamefont {Ma}}, \bibinfo
  {author} {\bibfnamefont {Z.}~\bibnamefont {Feng}}, \bibinfo {author}
  {\bibfnamefont {Y.}~\bibnamefont {Zhang}}, \bibinfo {author} {\bibfnamefont
  {L.}~\bibnamefont {Zhang}}, \bibinfo {author} {\bibfnamefont
  {H.}~\bibnamefont {Yang}}, \bibinfo {author} {\bibfnamefont {Y.}~\bibnamefont
  {Qi}}, \bibinfo {author} {\bibfnamefont {Z.~Y.}\ \bibnamefont {Meng}},
  \bibinfo {author} {\bibfnamefont {Y.-C.}\ \bibnamefont {Wang}}, \bibinfo
  {author} {\bibfnamefont {Y.}~\bibnamefont {Shi}},\ and\ \bibinfo {author}
  {\bibfnamefont {S.}~\bibnamefont {Li}},\ }\bibfield  {title} {\bibinfo
  {title} {Nonlocal effects of low-energy excitations in quantum-spin-liquid
  candidate cu$_3$zn(oh)$_6$fbr},\ }\href
  {https://doi.org/10.1088/0256-307X/38/9/097501} {\bibfield  {journal}
  {\bibinfo  {journal} {Chinese Physics Letters}\ }\textbf {\bibinfo {volume}
  {38}},\ \bibinfo {eid} {097501} (\bibinfo {year} {2021})}\BibitemShut
  {NoStop}%
\bibitem [{\citenamefont {Fu}\ \emph {et~al.}(2021)\citenamefont {Fu},
  \citenamefont {Lin}, \citenamefont {Wang}, \citenamefont {Liu}, \citenamefont
  {Huang}, \citenamefont {Jiang}, \citenamefont {Hao}, \citenamefont {Liu},
  \citenamefont {Zhang}, \citenamefont {Shi}, \citenamefont {Zhang},
  \citenamefont {Dai}, \citenamefont {Yu}, \citenamefont {Ye}, \citenamefont
  {Lee}, \citenamefont {Tan},\ and\ \citenamefont {Mei}}]{YingFu2021}%
  \BibitemOpen
  \bibfield  {author} {\bibinfo {author} {\bibfnamefont {Y.}~\bibnamefont
  {Fu}}, \bibinfo {author} {\bibfnamefont {M.-L.}\ \bibnamefont {Lin}},
  \bibinfo {author} {\bibfnamefont {L.}~\bibnamefont {Wang}}, \bibinfo {author}
  {\bibfnamefont {Q.}~\bibnamefont {Liu}}, \bibinfo {author} {\bibfnamefont
  {L.}~\bibnamefont {Huang}}, \bibinfo {author} {\bibfnamefont
  {W.}~\bibnamefont {Jiang}}, \bibinfo {author} {\bibfnamefont
  {Z.}~\bibnamefont {Hao}}, \bibinfo {author} {\bibfnamefont {C.}~\bibnamefont
  {Liu}}, \bibinfo {author} {\bibfnamefont {H.}~\bibnamefont {Zhang}}, \bibinfo
  {author} {\bibfnamefont {X.}~\bibnamefont {Shi}}, \bibinfo {author}
  {\bibfnamefont {J.}~\bibnamefont {Zhang}}, \bibinfo {author} {\bibfnamefont
  {J.}~\bibnamefont {Dai}}, \bibinfo {author} {\bibfnamefont {D.}~\bibnamefont
  {Yu}}, \bibinfo {author} {\bibfnamefont {F.}~\bibnamefont {Ye}}, \bibinfo
  {author} {\bibfnamefont {P.~A.}\ \bibnamefont {Lee}}, \bibinfo {author}
  {\bibfnamefont {P.-H.}\ \bibnamefont {Tan}},\ and\ \bibinfo {author}
  {\bibfnamefont {J.-W.}\ \bibnamefont {Mei}},\ }\bibfield  {title} {\bibinfo
  {title} {Dynamic fingerprint of fractionalized excitations in
  single-crystalline {Cu$_3$Zn(OH)$_6$FBr}},\ }\href
  {https://doi.org/10.1038/s41467-021-23381-9} {\bibfield  {journal} {\bibinfo
  {journal} {Nature Communications}\ }\textbf {\bibinfo {volume} {12}},\
  \bibinfo {pages} {3048} (\bibinfo {year} {2021})}\BibitemShut {NoStop}%
\bibitem [{\citenamefont {Zeng}\ \emph {et~al.}(2021)\citenamefont {Zeng},
  \citenamefont {Ma}, \citenamefont {Wu}, \citenamefont {Li}, \citenamefont
  {Tao}, \citenamefont {Lu}, \citenamefont {hui Chen}, \citenamefont {Mi},
  \citenamefont {Song}, \citenamefont {Cao}, \citenamefont {Che}, \citenamefont
  {Li}, \citenamefont {Li}, \citenamefont {Luo}, \citenamefont {Meng},\ and\
  \citenamefont {Li}}]{ZengZ21}%
  \BibitemOpen
  \bibfield  {author} {\bibinfo {author} {\bibfnamefont {Z.}~\bibnamefont
  {Zeng}}, \bibinfo {author} {\bibfnamefont {X.}~\bibnamefont {Ma}}, \bibinfo
  {author} {\bibfnamefont {S.}~\bibnamefont {Wu}}, \bibinfo {author}
  {\bibfnamefont {H.-F.}\ \bibnamefont {Li}}, \bibinfo {author} {\bibfnamefont
  {Z.}~\bibnamefont {Tao}}, \bibinfo {author} {\bibfnamefont {X.}~\bibnamefont
  {Lu}}, \bibinfo {author} {\bibfnamefont {X.}~\bibnamefont {hui Chen}},
  \bibinfo {author} {\bibfnamefont {J.-X.}\ \bibnamefont {Mi}}, \bibinfo
  {author} {\bibfnamefont {S.}~\bibnamefont {Song}}, \bibinfo {author}
  {\bibfnamefont {G.}~\bibnamefont {Cao}}, \bibinfo {author} {\bibfnamefont
  {G.}~\bibnamefont {Che}}, \bibinfo {author} {\bibfnamefont {K.}~\bibnamefont
  {Li}}, \bibinfo {author} {\bibfnamefont {G.}~\bibnamefont {Li}}, \bibinfo
  {author} {\bibfnamefont {H.}~\bibnamefont {Luo}}, \bibinfo {author}
  {\bibfnamefont {Z.~Y.}\ \bibnamefont {Meng}},\ and\ \bibinfo {author}
  {\bibfnamefont {S.}~\bibnamefont {Li}},\ }\href@noop {} {\bibinfo {title}
  {{Possible Dirac quantum spin liquid in a kagome quantum antiferromagnet
  YCu$_3$(OH)$_6$Br$_2$[Br$_x$(OH)$_{1-x}$]}}} (\bibinfo {year} {2021}),\
  \Eprint {https://arxiv.org/abs/arXiv:2107.11942} {arXiv:2107.11942}
  \BibitemShut {NoStop}%
\bibitem [{\citenamefont {Siddharthan}\ and\ \citenamefont
  {Georges}(2001)}]{SiddharthanR01}%
  \BibitemOpen
  \bibfield  {author} {\bibinfo {author} {\bibfnamefont {R.}~\bibnamefont
  {Siddharthan}}\ and\ \bibinfo {author} {\bibfnamefont {A.}~\bibnamefont
  {Georges}},\ }\bibfield  {title} {\bibinfo {title} {Square kagome quantum
  antiferromagnet and the eight-vertex model},\ }\href
  {https://doi.org/10.1103/PhysRevB.65.014417} {\bibfield  {journal} {\bibinfo
  {journal} {Phys. Rev. B}\ }\textbf {\bibinfo {volume} {65}},\ \bibinfo
  {pages} {014417} (\bibinfo {year} {2001})}\BibitemShut {NoStop}%
\bibitem [{\citenamefont {Richter}\ \emph {et~al.}(2009)\citenamefont
  {Richter}, \citenamefont {Schulenburg}, \citenamefont {Tomczak},\ and\
  \citenamefont {Schmalfu$\beta$}}]{RichterJ09}%
  \BibitemOpen
  \bibfield  {author} {\bibinfo {author} {\bibfnamefont {J.}~\bibnamefont
  {Richter}}, \bibinfo {author} {\bibfnamefont {J.}~\bibnamefont
  {Schulenburg}}, \bibinfo {author} {\bibfnamefont {P.}~\bibnamefont
  {Tomczak}},\ and\ \bibinfo {author} {\bibfnamefont {D.}~\bibnamefont
  {Schmalfu$\beta$}},\ }\bibfield  {title} {\bibinfo {title} {The heisenberg
  antiferromagnet on the square-kagome lattice},\ }\href
  {https://doi.org/10.5488/CMP.12.3.507} {\bibfield  {journal} {\bibinfo
  {journal} {Condens. Matter Phys.}\ }\textbf {\bibinfo {volume} {12}},\
  \bibinfo {pages} {507} (\bibinfo {year} {2009})}\BibitemShut {NoStop}%
\bibitem [{\citenamefont {Rousochatzakis}\ \emph {et~al.}(2013)\citenamefont
  {Rousochatzakis}, \citenamefont {Moessner},\ and\ \citenamefont {van~den
  Brink}}]{RousochatzakisI13}%
  \BibitemOpen
  \bibfield  {author} {\bibinfo {author} {\bibfnamefont {I.}~\bibnamefont
  {Rousochatzakis}}, \bibinfo {author} {\bibfnamefont {R.}~\bibnamefont
  {Moessner}},\ and\ \bibinfo {author} {\bibfnamefont {J.}~\bibnamefont
  {van~den Brink}},\ }\bibfield  {title} {\bibinfo {title} {Frustrated
  magnetism and resonating valence bond physics in two-dimensional kagome-like
  magnets},\ }\href {https://doi.org/10.1103/PhysRevB.88.195109} {\bibfield
  {journal} {\bibinfo  {journal} {Phys. Rev. B}\ }\textbf {\bibinfo {volume}
  {88}},\ \bibinfo {pages} {195109} (\bibinfo {year} {2013})}\BibitemShut
  {NoStop}%
\bibitem [{\citenamefont {Nakano}\ and\ \citenamefont
  {Sakai}(2013)}]{NakanoH13}%
  \BibitemOpen
  \bibfield  {author} {\bibinfo {author} {\bibfnamefont {H.}~\bibnamefont
  {Nakano}}\ and\ \bibinfo {author} {\bibfnamefont {T.}~\bibnamefont {Sakai}},\
  }\bibfield  {title} {\bibinfo {title} {The two-dimensional {S=1/2 Heisenberg}
  antiferromagnet on the shuriken lattice -- a lattice composed of
  vertex-sharing triangles},\ }\href {https://doi.org/10.7566/JPSJ.82.083709}
  {\bibfield  {journal} {\bibinfo  {journal} {J. Phys. Soc. Jpn.}\ }\textbf
  {\bibinfo {volume} {82}},\ \bibinfo {pages} {083709} (\bibinfo {year}
  {2013})}\BibitemShut {NoStop}%
\bibitem [{\citenamefont {Ralko}\ and\ \citenamefont
  {Rousochatzakis}(2015)}]{RalkoA15}%
  \BibitemOpen
  \bibfield  {author} {\bibinfo {author} {\bibfnamefont {A.}~\bibnamefont
  {Ralko}}\ and\ \bibinfo {author} {\bibfnamefont {I.}~\bibnamefont
  {Rousochatzakis}},\ }\bibfield  {title} {\bibinfo {title}
  {Resonating-valence-bond physics is not always governed by the shortest
  tunneling loops},\ }\href {https://doi.org/10.1103/PhysRevLett.115.167202}
  {\bibfield  {journal} {\bibinfo  {journal} {Phys. Rev. Lett.}\ }\textbf
  {\bibinfo {volume} {115}},\ \bibinfo {pages} {167202} (\bibinfo {year}
  {2015})}\BibitemShut {NoStop}%
\bibitem [{\citenamefont {Morita}\ and\ \citenamefont
  {Tohyama}(2018)}]{MoritaK18}%
  \BibitemOpen
  \bibfield  {author} {\bibinfo {author} {\bibfnamefont {K.}~\bibnamefont
  {Morita}}\ and\ \bibinfo {author} {\bibfnamefont {T.}~\bibnamefont
  {Tohyama}},\ }\bibfield  {title} {\bibinfo {title} {Magnetic phase diagrams
  and magnetization plateaus of the spin-1/2 antiferromagnetic {Heisenberg}
  model on a square-kagome lattice with three nonequivalent exchange
  interactions},\ }\href {https://doi.org/10.7566/JPSJ.87.043704} {\bibfield
  {journal} {\bibinfo  {journal} {J. Phys. Soc. Jpn.}\ }\textbf {\bibinfo
  {volume} {87}},\ \bibinfo {pages} {043704} (\bibinfo {year}
  {2018})}\BibitemShut {NoStop}%
\bibitem [{\citenamefont {Lugan}\ \emph {et~al.}(2019)\citenamefont {Lugan},
  \citenamefont {Jaubert},\ and\ \citenamefont {Ralko}}]{LuganT19}%
  \BibitemOpen
  \bibfield  {author} {\bibinfo {author} {\bibfnamefont {T.}~\bibnamefont
  {Lugan}}, \bibinfo {author} {\bibfnamefont {L.~D.~C.}\ \bibnamefont
  {Jaubert}},\ and\ \bibinfo {author} {\bibfnamefont {A.}~\bibnamefont
  {Ralko}},\ }\bibfield  {title} {\bibinfo {title} {Topological nematic spin
  liquid on the square kagome lattice},\ }\href
  {https://doi.org/10.1103/PhysRevResearch.1.033147} {\bibfield  {journal}
  {\bibinfo  {journal} {Phys. Rev. Research}\ }\textbf {\bibinfo {volume}
  {1}},\ \bibinfo {pages} {033147} (\bibinfo {year} {2019})}\BibitemShut
  {NoStop}%
\bibitem [{\citenamefont {Astrakhantsev}\ \emph {et~al.}(2021)\citenamefont
  {Astrakhantsev}, \citenamefont {Ferrari}, \citenamefont {Niggemann},
  \citenamefont {M\"uller}, \citenamefont {Chauhan}, \citenamefont
  {Kshetrimayum}, \citenamefont {Ghosh}, \citenamefont {Regnault},
  \citenamefont {Thomale}, \citenamefont {Reuther}, \citenamefont {Neupert},\
  and\ \citenamefont {Iqbal}}]{AstrakhantsevN21}%
  \BibitemOpen
  \bibfield  {author} {\bibinfo {author} {\bibfnamefont {N.}~\bibnamefont
  {Astrakhantsev}}, \bibinfo {author} {\bibfnamefont {F.}~\bibnamefont
  {Ferrari}}, \bibinfo {author} {\bibfnamefont {N.}~\bibnamefont {Niggemann}},
  \bibinfo {author} {\bibfnamefont {T.}~\bibnamefont {M\"uller}}, \bibinfo
  {author} {\bibfnamefont {A.}~\bibnamefont {Chauhan}}, \bibinfo {author}
  {\bibfnamefont {A.}~\bibnamefont {Kshetrimayum}}, \bibinfo {author}
  {\bibfnamefont {P.}~\bibnamefont {Ghosh}}, \bibinfo {author} {\bibfnamefont
  {N.}~\bibnamefont {Regnault}}, \bibinfo {author} {\bibfnamefont
  {R.}~\bibnamefont {Thomale}}, \bibinfo {author} {\bibfnamefont
  {J.}~\bibnamefont {Reuther}}, \bibinfo {author} {\bibfnamefont
  {T.}~\bibnamefont {Neupert}},\ and\ \bibinfo {author} {\bibfnamefont
  {Y.}~\bibnamefont {Iqbal}},\ }\bibfield  {title} {\bibinfo {title} {Pinwheel
  valence bond crystal ground state of the spin-$\frac{1}{2}$ {Heisenberg}
  antiferromagnet on the shuriken lattice},\ }\href
  {https://doi.org/10.1103/PhysRevB.104.L220408} {\bibfield  {journal}
  {\bibinfo  {journal} {Phys. Rev. B}\ }\textbf {\bibinfo {volume} {104}},\
  \bibinfo {pages} {L220408} (\bibinfo {year} {2021})}\BibitemShut {NoStop}%
\bibitem [{\citenamefont {Richter}\ \emph {et~al.}(2022)\citenamefont
  {Richter}, \citenamefont {Derzhko},\ and\ \citenamefont
  {Schnack}}]{RichterJ22}%
  \BibitemOpen
  \bibfield  {author} {\bibinfo {author} {\bibfnamefont {J.}~\bibnamefont
  {Richter}}, \bibinfo {author} {\bibfnamefont {O.}~\bibnamefont {Derzhko}},\
  and\ \bibinfo {author} {\bibfnamefont {J.}~\bibnamefont {Schnack}},\
  }\href@noop {} {\bibinfo {title} {Thermodynamics of the spin-half
  square-kagome lattice antiferromagnet}} (\bibinfo {year} {2022}),\ \Eprint
  {https://arxiv.org/abs/arXiv:2202.07357} {arXiv:2202.07357} \BibitemShut
  {NoStop}%
\bibitem [{\citenamefont {Fujihala}\ \emph {et~al.}(2020)\citenamefont
  {Fujihala}, \citenamefont {Morita}, \citenamefont {Mole}, \citenamefont
  {Mitsuda}, \citenamefont {Tohyama}, \citenamefont {ichiro Yano},
  \citenamefont {Yu}, \citenamefont {Sota}, \citenamefont {Kuwai},
  \citenamefont {Koda}, \citenamefont {Okabe}, \citenamefont {Lee},
  \citenamefont {Itoh}, \citenamefont {Hawai}, \citenamefont {Masuda},
  \citenamefont {Sagayama}, \citenamefont {Matsuo}, \citenamefont {Kindo},
  \citenamefont {Ohira-Kawamura},\ and\ \citenamefont
  {Nakajima}}]{FujihalaM20}%
  \BibitemOpen
  \bibfield  {author} {\bibinfo {author} {\bibfnamefont {M.}~\bibnamefont
  {Fujihala}}, \bibinfo {author} {\bibfnamefont {K.}~\bibnamefont {Morita}},
  \bibinfo {author} {\bibfnamefont {R.}~\bibnamefont {Mole}}, \bibinfo {author}
  {\bibfnamefont {S.}~\bibnamefont {Mitsuda}}, \bibinfo {author} {\bibfnamefont
  {T.}~\bibnamefont {Tohyama}}, \bibinfo {author} {\bibfnamefont
  {S.}~\bibnamefont {ichiro Yano}}, \bibinfo {author} {\bibfnamefont
  {D.}~\bibnamefont {Yu}}, \bibinfo {author} {\bibfnamefont {S.}~\bibnamefont
  {Sota}}, \bibinfo {author} {\bibfnamefont {T.}~\bibnamefont {Kuwai}},
  \bibinfo {author} {\bibfnamefont {A.}~\bibnamefont {Koda}}, \bibinfo {author}
  {\bibfnamefont {H.}~\bibnamefont {Okabe}}, \bibinfo {author} {\bibfnamefont
  {H.}~\bibnamefont {Lee}}, \bibinfo {author} {\bibfnamefont {S.}~\bibnamefont
  {Itoh}}, \bibinfo {author} {\bibfnamefont {T.}~\bibnamefont {Hawai}},
  \bibinfo {author} {\bibfnamefont {T.}~\bibnamefont {Masuda}}, \bibinfo
  {author} {\bibfnamefont {H.}~\bibnamefont {Sagayama}}, \bibinfo {author}
  {\bibfnamefont {A.}~\bibnamefont {Matsuo}}, \bibinfo {author} {\bibfnamefont
  {K.}~\bibnamefont {Kindo}}, \bibinfo {author} {\bibfnamefont
  {S.}~\bibnamefont {Ohira-Kawamura}},\ and\ \bibinfo {author} {\bibfnamefont
  {K.}~\bibnamefont {Nakajima}},\ }\bibfield  {title} {\bibinfo {title}
  {Gapless spin liquid in a square-kagome lattice antiferromagnet},\ }\href
  {https://doi.org/10.1038/s41467-020-17235-z} {\bibfield  {journal} {\bibinfo
  {journal} {Nat. Commun.}\ }\textbf {\bibinfo {volume} {11}},\ \bibinfo
  {pages} {3429} (\bibinfo {year} {2020})}\BibitemShut {NoStop}%
\bibitem [{\citenamefont {Yakubovich}\ \emph {et~al.}(2021)\citenamefont
  {Yakubovich}, \citenamefont {Shvanskaya}, \citenamefont {Kiriukhina},
  \citenamefont {Volkov}, \citenamefont {Dimitrova},\ and\ \citenamefont
  {Vasiliev}}]{YakubovichOV21}%
  \BibitemOpen
  \bibfield  {author} {\bibinfo {author} {\bibfnamefont {O.~V.}\ \bibnamefont
  {Yakubovich}}, \bibinfo {author} {\bibfnamefont {L.~V.}\ \bibnamefont
  {Shvanskaya}}, \bibinfo {author} {\bibfnamefont {G.~V.}\ \bibnamefont
  {Kiriukhina}}, \bibinfo {author} {\bibfnamefont {A.~S.}\ \bibnamefont
  {Volkov}}, \bibinfo {author} {\bibfnamefont {O.~V.}\ \bibnamefont
  {Dimitrova}},\ and\ \bibinfo {author} {\bibfnamefont {A.~N.}\ \bibnamefont
  {Vasiliev}},\ }\bibfield  {title} {\bibinfo {title} {Hydrothermal synthesis
  and a composite crystal structure of
  {Na$_6$Cu$_7$BiO$_4$(PO$_4$)$_4$[Cl,(OH)]$_3$} as a candidate for quantum
  spin liquid},\ }\href {https://doi.org/10.1021/acs.inorgchem.1c01459}
  {\bibfield  {journal} {\bibinfo  {journal} {Inorg. Chem.}\ }\textbf {\bibinfo
  {volume} {60}},\ \bibinfo {pages} {11450} (\bibinfo {year}
  {2021})}\BibitemShut {NoStop}%
\bibitem [{\citenamefont {de~Vries}\ \emph {et~al.}(2008)\citenamefont
  {de~Vries}, \citenamefont {Kamenev}, \citenamefont {Kockelmann},
  \citenamefont {Sanchez-Benitez},\ and\ \citenamefont {Harrison}}]{VriesMA08}%
  \BibitemOpen
  \bibfield  {author} {\bibinfo {author} {\bibfnamefont {M.~A.}\ \bibnamefont
  {de~Vries}}, \bibinfo {author} {\bibfnamefont {K.~V.}\ \bibnamefont
  {Kamenev}}, \bibinfo {author} {\bibfnamefont {W.~A.}\ \bibnamefont
  {Kockelmann}}, \bibinfo {author} {\bibfnamefont {J.}~\bibnamefont
  {Sanchez-Benitez}},\ and\ \bibinfo {author} {\bibfnamefont {A.}~\bibnamefont
  {Harrison}},\ }\bibfield  {title} {\bibinfo {title} {Magnetic ground state of
  an experimental {$S=1/2$} kagome antiferromagnet},\ }\href
  {https://doi.org/10.1103/PhysRevLett.100.157205} {\bibfield  {journal}
  {\bibinfo  {journal} {Phys. Rev. Lett.}\ }\textbf {\bibinfo {volume} {100}},\
  \bibinfo {pages} {157205} (\bibinfo {year} {2008})}\BibitemShut {NoStop}%
\bibitem [{\citenamefont {Freedman}\ \emph {et~al.}(2010)\citenamefont
  {Freedman}, \citenamefont {Han}, \citenamefont {Prodi}, \citenamefont
  {M{\"{u}}ller}, \citenamefont {Huang}, \citenamefont {Chen}, \citenamefont
  {Webb}, \citenamefont {Lee}, \citenamefont {McQueen},\ and\ \citenamefont
  {Nocera}}]{FreedmanDE10}%
  \BibitemOpen
  \bibfield  {author} {\bibinfo {author} {\bibfnamefont {D.~E.}\ \bibnamefont
  {Freedman}}, \bibinfo {author} {\bibfnamefont {T.~H.}\ \bibnamefont {Han}},
  \bibinfo {author} {\bibfnamefont {A.}~\bibnamefont {Prodi}}, \bibinfo
  {author} {\bibfnamefont {P.}~\bibnamefont {M{\"{u}}ller}}, \bibinfo {author}
  {\bibfnamefont {Q.-Z.}\ \bibnamefont {Huang}}, \bibinfo {author}
  {\bibfnamefont {Y.-S.}\ \bibnamefont {Chen}}, \bibinfo {author}
  {\bibfnamefont {S.~M.}\ \bibnamefont {Webb}}, \bibinfo {author}
  {\bibfnamefont {Y.~S.}\ \bibnamefont {Lee}}, \bibinfo {author} {\bibfnamefont
  {T.~M.}\ \bibnamefont {McQueen}},\ and\ \bibinfo {author} {\bibfnamefont
  {D.~G.}\ \bibnamefont {Nocera}},\ }\bibfield  {title} {\bibinfo {title} {Site
  specific x-ray anomalous dispersion of the geometrically frustrated
  kagom\'{e} magnet, herbertsmithite, {ZnCu$_3$(OH)$_6$Cl$_2$}},\ }\href
  {https://doi.org/10.1021/ja1070398} {\bibfield  {journal} {\bibinfo
  {journal} {J. Am. Chem. Soc.}\ ,\ \bibinfo {pages} {16185}} (\bibinfo {year}
  {2010})}\BibitemShut {NoStop}%
\bibitem [{\citenamefont {{Huang}}\ \emph {et~al.}(2021)\citenamefont
  {{Huang}}, \citenamefont {{Xu}}, \citenamefont {{Wang}}, \citenamefont
  {{Zhao}}, \citenamefont {{Tu}}, \citenamefont {{Ni}}, \citenamefont {{Wang}},
  \citenamefont {{Pan}}, \citenamefont {{Fu}}, \citenamefont {{Hao}},
  \citenamefont {{Liu}}, \citenamefont {{Mei}},\ and\ \citenamefont
  {{Li}}}]{YYHuang2021}%
  \BibitemOpen
  \bibfield  {author} {\bibinfo {author} {\bibfnamefont {Y.~Y.}\ \bibnamefont
  {{Huang}}}, \bibinfo {author} {\bibfnamefont {Y.}~\bibnamefont {{Xu}}},
  \bibinfo {author} {\bibfnamefont {L.}~\bibnamefont {{Wang}}}, \bibinfo
  {author} {\bibfnamefont {C.~C.}\ \bibnamefont {{Zhao}}}, \bibinfo {author}
  {\bibfnamefont {C.~P.}\ \bibnamefont {{Tu}}}, \bibinfo {author}
  {\bibfnamefont {J.~M.}\ \bibnamefont {{Ni}}}, \bibinfo {author}
  {\bibfnamefont {L.~S.}\ \bibnamefont {{Wang}}}, \bibinfo {author}
  {\bibfnamefont {B.~L.}\ \bibnamefont {{Pan}}}, \bibinfo {author}
  {\bibfnamefont {Y.}~\bibnamefont {{Fu}}}, \bibinfo {author} {\bibfnamefont
  {Z.}~\bibnamefont {{Hao}}}, \bibinfo {author} {\bibfnamefont
  {C.}~\bibnamefont {{Liu}}}, \bibinfo {author} {\bibfnamefont {J.-W.}\
  \bibnamefont {{Mei}}},\ and\ \bibinfo {author} {\bibfnamefont {S.~Y.}\
  \bibnamefont {{Li}}},\ }\bibfield  {title} {\bibinfo {title} {{Heat Transport
  in Herbertsmithite: Can a Quantum Spin Liquid Survive Disorder?}},\
  }\href@noop {} {\bibfield  {journal} {\bibinfo  {journal} {arXiv e-prints}\
  ,\ \bibinfo {eid} {arXiv:2105.14749}} (\bibinfo {year} {2021})},\ \Eprint
  {https://arxiv.org/abs/2105.14749} {arXiv:2105.14749 [cond-mat.str-el]}
  \BibitemShut {NoStop}%
\bibitem [{\citenamefont {Kermarrec}\ \emph {et~al.}(2011)\citenamefont
  {Kermarrec}, \citenamefont {Mendels}, \citenamefont {Bert}, \citenamefont
  {Colman}, \citenamefont {Wills}, \citenamefont {Strobel}, \citenamefont
  {Bonville}, \citenamefont {Hillier},\ and\ \citenamefont
  {Amato}}]{KermarrecE11}%
  \BibitemOpen
  \bibfield  {author} {\bibinfo {author} {\bibfnamefont {E.}~\bibnamefont
  {Kermarrec}}, \bibinfo {author} {\bibfnamefont {P.}~\bibnamefont {Mendels}},
  \bibinfo {author} {\bibfnamefont {F.}~\bibnamefont {Bert}}, \bibinfo {author}
  {\bibfnamefont {R.~H.}\ \bibnamefont {Colman}}, \bibinfo {author}
  {\bibfnamefont {A.~S.}\ \bibnamefont {Wills}}, \bibinfo {author}
  {\bibfnamefont {P.}~\bibnamefont {Strobel}}, \bibinfo {author} {\bibfnamefont
  {P.}~\bibnamefont {Bonville}}, \bibinfo {author} {\bibfnamefont
  {A.}~\bibnamefont {Hillier}},\ and\ \bibinfo {author} {\bibfnamefont
  {A.}~\bibnamefont {Amato}},\ }\bibfield  {title} {\bibinfo {title}
  {Spin-liquid ground state in the frustrated kagome antiferromagnet
  {MgCu${}_{3}$(OH)${}_{6}$Cl${}_{2}$}},\ }\href
  {https://doi.org/10.1103/PhysRevB.84.100401} {\bibfield  {journal} {\bibinfo
  {journal} {Phys. Rev. B}\ }\textbf {\bibinfo {volume} {84}},\ \bibinfo
  {pages} {100401} (\bibinfo {year} {2011})}\BibitemShut {NoStop}%
\bibitem [{\citenamefont {Li}\ \emph {et~al.}(2014)\citenamefont {Li},
  \citenamefont {Pan}, \citenamefont {Li}, \citenamefont {Tong}, \citenamefont
  {Ling}, \citenamefont {Yang}, \citenamefont {Wang}, \citenamefont {Chen},
  \citenamefont {Wu},\ and\ \citenamefont {Zhang}}]{LIY14}%
  \BibitemOpen
  \bibfield  {author} {\bibinfo {author} {\bibfnamefont {Y.}~\bibnamefont
  {Li}}, \bibinfo {author} {\bibfnamefont {B.}~\bibnamefont {Pan}}, \bibinfo
  {author} {\bibfnamefont {S.}~\bibnamefont {Li}}, \bibinfo {author}
  {\bibfnamefont {W.}~\bibnamefont {Tong}}, \bibinfo {author} {\bibfnamefont
  {L.}~\bibnamefont {Ling}}, \bibinfo {author} {\bibfnamefont {Z.}~\bibnamefont
  {Yang}}, \bibinfo {author} {\bibfnamefont {J.}~\bibnamefont {Wang}}, \bibinfo
  {author} {\bibfnamefont {Z.}~\bibnamefont {Chen}}, \bibinfo {author}
  {\bibfnamefont {Z.}~\bibnamefont {Wu}},\ and\ \bibinfo {author}
  {\bibfnamefont {Q.}~\bibnamefont {Zhang}},\ }\bibfield  {title} {\bibinfo
  {title} {Gapless quantum spin liquid in the {$S$} = 1/2 anisotropic kagome
  antiferromagnet {ZnCu$_3$(OH)$_6$SO$_4$}},\ }\href@noop {} {\bibfield
  {journal} {\bibinfo  {journal} {New J. Phys.}\ }\textbf {\bibinfo {volume}
  {16}},\ \bibinfo {pages} {093011} (\bibinfo {year} {2014})}\BibitemShut
  {NoStop}%
\bibitem [{\citenamefont {Ran}\ \emph {et~al.}(2007)\citenamefont {Ran},
  \citenamefont {Hermele}, \citenamefont {Lee},\ and\ \citenamefont
  {Wen}}]{RanY07}%
  \BibitemOpen
  \bibfield  {author} {\bibinfo {author} {\bibfnamefont {Y.}~\bibnamefont
  {Ran}}, \bibinfo {author} {\bibfnamefont {M.}~\bibnamefont {Hermele}},
  \bibinfo {author} {\bibfnamefont {P.~A.}\ \bibnamefont {Lee}},\ and\ \bibinfo
  {author} {\bibfnamefont {X.-G.}\ \bibnamefont {Wen}},\ }\bibfield  {title}
  {\bibinfo {title} {Projected-wave-function study of the spin-$1/2$ heisenberg
  model on the kagom\'e lattice},\ }\href
  {https://doi.org/10.1103/PhysRevLett.98.117205} {\bibfield  {journal}
  {\bibinfo  {journal} {Phys. Rev. Lett.}\ }\textbf {\bibinfo {volume} {98}},\
  \bibinfo {pages} {117205} (\bibinfo {year} {2007})}\BibitemShut {NoStop}%
\bibitem [{\citenamefont {Xu}\ \emph {et~al.}(2019)\citenamefont {Xu},
  \citenamefont {Qi}, \citenamefont {Zhang}, \citenamefont {Assaad},
  \citenamefont {Xu},\ and\ \citenamefont {Meng}}]{XuXY19}%
  \BibitemOpen
  \bibfield  {author} {\bibinfo {author} {\bibfnamefont {X.~Y.}\ \bibnamefont
  {Xu}}, \bibinfo {author} {\bibfnamefont {Y.}~\bibnamefont {Qi}}, \bibinfo
  {author} {\bibfnamefont {L.}~\bibnamefont {Zhang}}, \bibinfo {author}
  {\bibfnamefont {F.~F.}\ \bibnamefont {Assaad}}, \bibinfo {author}
  {\bibfnamefont {C.}~\bibnamefont {Xu}},\ and\ \bibinfo {author}
  {\bibfnamefont {Z.~Y.}\ \bibnamefont {Meng}},\ }\bibfield  {title} {\bibinfo
  {title} {Monte carlo study of lattice compact quantum electrodynamics with
  {Fermionic} matter: {The} parent state of quantum phases},\ }\href
  {https://doi.org/10.1103/PhysRevX.9.021022} {\bibfield  {journal} {\bibinfo
  {journal} {Phys. Rev. X}\ }\textbf {\bibinfo {volume} {9}},\ \bibinfo {pages}
  {021022} (\bibinfo {year} {2019})}\BibitemShut {NoStop}%
\bibitem [{\citenamefont {Takahashi}(1989)}]{TakahashiM89}%
  \BibitemOpen
  \bibfield  {author} {\bibinfo {author} {\bibfnamefont {M.}~\bibnamefont
  {Takahashi}},\ }\bibfield  {title} {\bibinfo {title} {Modified spin-wave
  theory of a square-lattice antiferromagnet},\ }\href
  {https://doi.org/10.1103/PhysRevB.40.2494} {\bibfield  {journal} {\bibinfo
  {journal} {Phys. Rev. B}\ }\textbf {\bibinfo {volume} {40}},\ \bibinfo
  {pages} {2494} (\bibinfo {year} {1989})}\BibitemShut {NoStop}%
\bibitem [{\citenamefont {Liu}(1990)}]{LiuBG90}%
  \BibitemOpen
  \bibfield  {author} {\bibinfo {author} {\bibfnamefont {B.-G.}\ \bibnamefont
  {Liu}},\ }\bibfield  {title} {\bibinfo {title} {{Low-temperature properties
  of the quasi-two-dimensional antiferromagnetic Heisenberg model}},\ }\href
  {https://doi.org/10.1103/PhysRevB.41.9563} {\bibfield  {journal} {\bibinfo
  {journal} {Phys. Rev. B}\ }\textbf {\bibinfo {volume} {41}},\ \bibinfo
  {pages} {9563} (\bibinfo {year} {1990})}\BibitemShut {NoStop}%
\bibitem [{\citenamefont {Shimizu}\ \emph {et~al.}(2003)\citenamefont
  {Shimizu}, \citenamefont {Matsumoto}, \citenamefont {Goto}, \citenamefont
  {Chandrasekhar~Rao}, \citenamefont {Yoshimura},\ and\ \citenamefont
  {Kosuge}}]{ShimizuT03}%
  \BibitemOpen
  \bibfield  {author} {\bibinfo {author} {\bibfnamefont {T.}~\bibnamefont
  {Shimizu}}, \bibinfo {author} {\bibfnamefont {T.}~\bibnamefont {Matsumoto}},
  \bibinfo {author} {\bibfnamefont {A.}~\bibnamefont {Goto}}, \bibinfo {author}
  {\bibfnamefont {T.~V.}\ \bibnamefont {Chandrasekhar~Rao}}, \bibinfo {author}
  {\bibfnamefont {K.}~\bibnamefont {Yoshimura}},\ and\ \bibinfo {author}
  {\bibfnamefont {K.}~\bibnamefont {Kosuge}},\ }\bibfield  {title} {\bibinfo
  {title} {{Spin susceptibility and superexchange interaction in the
  antiferromagnet CuO}},\ }\href {https://doi.org/10.1103/PhysRevB.68.224433}
  {\bibfield  {journal} {\bibinfo  {journal} {Phys. Rev. B}\ }\textbf {\bibinfo
  {volume} {68}},\ \bibinfo {pages} {224433} (\bibinfo {year}
  {2003})}\BibitemShut {NoStop}%
\bibitem [{\citenamefont {Rocquefelte}\ \emph {et~al.}(2012)\citenamefont
  {Rocquefelte}, \citenamefont {Schwarz},\ and\ \citenamefont
  {Blaha}}]{RocquefelteX12}%
  \BibitemOpen
  \bibfield  {author} {\bibinfo {author} {\bibfnamefont {X.}~\bibnamefont
  {Rocquefelte}}, \bibinfo {author} {\bibfnamefont {K.}~\bibnamefont
  {Schwarz}},\ and\ \bibinfo {author} {\bibfnamefont {P.}~\bibnamefont
  {Blaha}},\ }\bibfield  {title} {\bibinfo {title} {Theoretical investigation
  of the magnetic exchange interactions in copper({II}) oxides under chemical
  and physical pressures},\ }\href
  {https://doi.org/https://doi.org/10.1038/srep00759} {\bibfield  {journal}
  {\bibinfo  {journal} {Sci. Rep.}\ }\textbf {\bibinfo {volume} {2}},\ \bibinfo
  {pages} {759} (\bibinfo {year} {2012})}\BibitemShut {NoStop}%
\end{thebibliography}
\end{document}